\documentclass[pra,twocolumn,amssymb,superscriptaddress,showpacs,aps,floatfix]{revtex4}
\newcommand{\be}{\begin{equation}}
\newcommand{\ee}{\end{equation}}
\newcommand{\bea}{\begin{eqnarray}}
\newcommand{\eea}{\end{eqnarray}}
\usepackage{graphicx}
\usepackage{epsfig}

\begin{document}

\title{Fractional Quantum Hall states in the vicinity of Mott plateaus}

\author{R.~O.~Umucal{\i}lar}
\email{onur@fen.bilkent.edu.tr}
\affiliation{Department of Physics, Bilkent University, 06800 Ankara, Turkey}
\affiliation{Laboratory for Atomic and Solid State Physics, Cornell University, Ithaca, NY 14853, USA}
\author{Erich J. Mueller}
\affiliation{Laboratory for Atomic and Solid State Physics, Cornell University, Ithaca, NY 14853, USA}

\date{\today}

\begin{abstract}
We perform variational Monte-Carlo calculations to show that bosons in a rotating optical lattice will form analogs of fractional quantum Hall states when the tunneling is sufficiently weak compared to the interactions and the deviation of density from an integer is commensurate with the effective magnetic field.  We compare the energies of superfluid and correlated states to one-another and to the energies found in full configuration-interaction calculations on small systems.  We look at overlaps between our variational states and the exact ground-state, characterizing the ways in which fractional quantum Hall effect correlations manifest themselves near the Mott insulating state.  We explore the experimental signatures of these states.
\end{abstract}


\maketitle
\section{Introduction}
We consider the interplay between three paradigmatic quantum states of bosons in rotating lattices: Mott insulators, superfluids, and fractional quantum Hall states.
The Mott insulator is found when there is an integer number of particles per lattice site, and the tunneling is sufficiently suppressed relative to the interactions.  It is an incompressible state, where interactions freeze the particles in place.  In the standard cartoon, when the density of such a system is tuned away from commensurability, the excess particles (or holes) ``skate" across the frozen Mott sea, forming a superfluid.  If the system is rotating, one expects that the collective motion of this superfluid will produce a vortex lattice.  In 2007, Umucal{\i}lar and Oktel \cite{Umucalilar} argued that when the rotation rate is high enough that the number of vortices is comparable to the number of excess particles, then the superfluid will be unstable to forming a correlated state of matter with particles bound to vortices -- a situation analogous to that found in the fractional quantum Hall state.  They supported this argument by estimating the energy of the superfluid and the correlated state.  Here we confirm this scenario through more rigorous calculations.  By using Monte-Carlo techniques we compare the energy of variational states describing fractional quantum Hall states and superfluid vortex lattices with each-other.  We also compare these energies with exact results calculated for small numbers of particles.  We find that there is a range of parameters for which the fractional quantum Hall states are more favorable than superfluid states.  We note, however, that the energy differences between these states scales as the tunelling energy, and can be quite small.

Most previous studies of analogs of fractional quantum Hall states in optical lattices have focussed on the low density limit, where there are much fewer than one particle per site.  In the context of cold atoms, Hafezi, S{\o}rensen, Demler and Lukin \cite{Hafezi} gave an excellent review of the basic physics of this limit (including symmetry and topology arguments), and argued that one can continuously deform a Mott insulating state into a fractional quantum Hall state by varying the strength of an additional superlattice potential \cite{Sorensen}.  They also proposed using Bragg spectroscopy to probe these states.  Palmer, Klein, and Jaksch \cite{Palmer} performed a number of calculations focussed on the role of the trap, detection schemes, and on inhomogeneities which can spontaneously appear in these systems.
Bhat, et al.  \cite{Bhat} carried out full configuration-interaction calculations for a small number of particles in a rotating lattice with hard-wall boundary conditions.
M\"oller and Cooper analyzed the relevance of composite fermion wavefunctions to describing these systems \cite{Moller}.
Nigel Cooper recently produced a review of the physics of rotating cold atom clouds including analogs of the quantum Hall effect in lattices \cite{Cooper}.
These, and our present study, build on initial works motivated by solid state systems \cite{solidstate}.

Translating these arguments to higher densities is not completely
trivial.  The superfluid near the Mott state is more complicated
than the standard cartoon suggests.  For example, the mean-field
description treats it as a two-component plasma of particles and
holes, with a small imbalance between the density of particles and
holes. Despite these complications, we find that when the
deviation of the particle density from an integer value is
commensurate with the magnetic flux one can indeed see analogs of
the fractional quantum Hall effect.

We start our analysis with the well-known Bose-Hubbard Hamiltonian
in an effective magnetic field \bea \label{Bose-Hubbard} H_0 =
-t\sum_{\langle i,j\rangle}a_i^\dag a_j e^{i
A_{ij}}+\frac{U}{2}\sum_i \hat{n}_i (\hat{n}_i-1), \eea where
$a_i$ ($a_i^\dag$) is the bosonic annihilation (creation) operator
at site $i$ and $\hat{n}_i = a_i^\dag a_i$ is the number operator.
The tunneling is parameterized by $t$ and on-site interactions by
$U$. We use the Landau gauge $\mathbf{A} = (-By,0)$, so the phases
$A_{ij} = \exp(ie/\hbar c
\int^{\mathbf{r}_i}_{\mathbf{r}_j}\mathbf{A}\cdot d\mathbf{l})$
acquired when hopping in $\pm$ $x$-direction are $\mp 2 \pi \alpha
i_y$, where $i_y $ is the $y$ coordinate scaled by lattice
constant $a$, and in $y$-direction $A_{ij} = 0$. Here, $\alpha =
Ba^2/(hc/e) = p/q$ is the flux quantum per plaquette and we take
$p$ and $q$ to be relatively prime integers. The single particle
spectrum for this problem is the famous Hofstadter butterfly
\cite{Hofstadter}. The phase boundary between the Mott insulator
and superfluid carries signatures of this single-particle physics
\cite{Niemeyer,Oktel,Umucalilar,Goldbaum}.  Away from the tips of
the Mott lobes, the physics of the superfluid-Mott transition of
the non-rotating system is in the universality class of the dilute
Bose gas.  Thus we expect that phenomena which can be seen in the
dilute Bose gas will occur there, including the analogs of
fractional quantum Hall physics which we are exploring here.

\section{Variational Wavefunction}
\subsection{Laughlin State}\label{variational}
We consider the variational ansatz
\be\label{FQH ansatz}
|\Psi\rangle = \sum_{z_1,...,z_N}\psi(z_1,...,z_N)a^{\dag}_{z_1}...a^{\dag}_{z_N}|\Psi_{MI}\rangle,
\ee
where $|\Psi_{MI}\rangle=\prod_j (a_j^\dagger)^{n_0}/\sqrt{n_0!} |{\rm vac}\rangle$ is the Mott insulator state with $n_0$ particles per site and $\psi$ is the Laughlin wavefunction \cite{Laughlin} with filling $\nu=1/m$.  To describe bosons, $m$ must be even.  The complex coordinate $z_i = x_i+i y_i$ specifies the location of the $i$th particle, with $i$ running from 1 to $N$, where $N$ is the number of excess particles.  The sum over $z_i$ is a sum over all lattice sites.  To describe a state with excess holes, we replace $a^\dag$ with $a$.

To minimize the role of boundaries, the model in (\ref{Bose-Hubbard}) is typically either solved on a sphere or a torus \cite{solidstate,Haldane}.  We will work in an $L\times L$ torus geometry, corresponding to quasiperiodic boundary conditions
\bea\label{boundary conditions}
\psi(...,z_k+L,...)&=&\psi(...,z_k,...) \\
\psi(...,z_k+iL,...)&=& e^{-i\frac{2\pi m N}{L}x_k}\psi(...,z_k,...), \nonumber
\eea
For these boundary conditions the Laughlin wavefunction can explicitly be written as
\cite{Haldane}
\begin{widetext}
\bea\label{Wavefunction}
\psi(z_1,...,z_N) = \mathcal{N}e^{iK_x\sum_i x_i}e^{-K_y\sum_i y_i}e^{-\frac{\pi m N}{L^2}\sum_i y_i^2}\prod_{\beta=1}^m \vartheta_1\big[(Z-Z_{\beta})\frac{\pi}{L}\big]\prod_{i<j}^{N}\Big\{\vartheta_1\big[(z_i-z_j)\frac{\pi}{L}\big]\Big\}^m.
\eea
\end{widetext}
Here, $\mathcal{N}$ is the normalization factor, $Z = \sum_{i}z_i$ is $N$ times the center-of-mass coordinate, $Z_{\beta} = X_{\beta}+i Y_{\beta}$ are the {\em a-priori} arbitrary locations of the center-of-mass zeros.  To satisfy the boundary conditions, one requires $\sum_{\beta}X_\beta = n_1 L$ ($n_1\in \mathbb{Z}$), $K_x = 2\pi n_2/L$ ($n_2\in \mathbb{Z}$), and $K_y = -2\pi\sum_{\beta}Y_\beta/L^2$.
The quasi-periodic Jacobi theta functions are defined by
\bea\label{Theta function}
\vartheta_1(z,e^{i\pi \tau}) = \sum_{-\infty}^{\infty}(-1)^{n-1/2}e^{i\pi\tau (n+1/2)^2}e^{(2n+1)iz}. \nonumber
\eea
For our square geometry $\tau = i$. This function is odd with respect to $z$ and has the following quasi-periodicity properties: $\vartheta_1(z+\pi)=-\vartheta_1(z)$ and $\vartheta_1(z+\tau \pi)=-e^{-i\pi\tau} e^{-2iz}\vartheta_1(z)$. The relation between the flux quantum per plaquette $\alpha = N_{\phi}/L^2$, filling fraction $\nu = N/N_{\phi}$, and excess particle density $\varepsilon = N/L^2$ is succinctly given by  $\alpha \nu = \varepsilon$, where $N_{\phi}$ denotes the number of flux quanta in the $L\times L$ lattice we consider. In what follows, we will restrict ourselves to the $\nu = 1/2$ Laughlin state ($m = 2$), so that the commensurability requirement between the magnetic flux and particle density becomes $\alpha = 2\varepsilon$.

\subsection{Superfluid State}
We will compare the Laughlin state introduced in Sec.~\ref{variational} with a  Gutzwiller mean field state
\begin{equation}\label{gutz}
|\Psi_{MF}\rangle = \prod_i\left(\sum_n f^i_n |n\rangle_i\right),
\end{equation}
where $f^i_n$ are variational parameters.
This wavefunction is commonly used to describe the superfluid in the Bose-Hubbard model \cite{Bloch}.  It is exact in the non-interacting limit and captures the effect of number squeezing.  Its main deficit is that it does not capture any of the short-range correlations in the superfluid.  Regardless, the energies it produces are good estimates of the superfluid energy.
In the non-rotating case, the superfluid is translationally invariant, and the coefficients $f_n^i$ are independent of $i$.  In our case, where the lattice is rotating, a vortex lattice forms, breaking translational invariance.

Near the Mott lobe, the site occupations are dominated by $n=n_0$ and $n=n_0\pm1$: that is it is extremely unlikely to have more than one extra particle or hole on a given site.  We therefore truncate our basis to only these three values of $n$.  This will also facilitate direct comparison with configuration-interaction calculations using the same truncated basis. We work in an $L\times L$ lattice, using the boundary conditions which are equivalent to those in Eq.~(\ref{boundary conditions}).

The numerical techniques for optimizing the $f_n^{i}$ are well documented \cite{Goldbaum}, and we will not repeat the detailed discussion here.  These can be described in terms of a variational calculation where one minimizes $\langle \Psi_{MF}|H_0 |\Psi_{MF}\rangle$ with the constraints that the total number of particles $M$ and normalization $\langle\Psi_{MF}|\Psi_{MF}\rangle$ are fixed: this involves introducing the chemical potential $\mu$ and a number of other Lagrange multipliers.  In practice it is more convenient to write $H=H_0-\mu M$, and follow an iterative procedure based upon mean field theory.  These two approaches are completely equivalent.  In comparing energies with our other variational state, one must be cautious and be sure to use $\langle H_0\rangle=\langle H\rangle+\mu M$.

\section{Exact results on small systems}\label{exact}

\subsection{Approach and Results}

For small systems we can exactly diagonalize the Hamiltonian in Eq.~(\ref{Bose-Hubbard}), taking a configuration-interaction approach where we truncate the allowed number of particles on a given site to be $n_0$, $n_0-1$, or $n_0+1$.  For definiteness we take $n_0=1$: changing this value just scales the hopping matrix elements $t$.  For these small system sizes we can also directly calculate $\langle\Psi|H_0|\Psi\rangle$. In Sec.~\ref{monte-carlo} we will discuss larger systems where we need to resort to a Monte-Carlo algorithm for calculating this energy.

\begin{figure}
\includegraphics[scale=.42]
{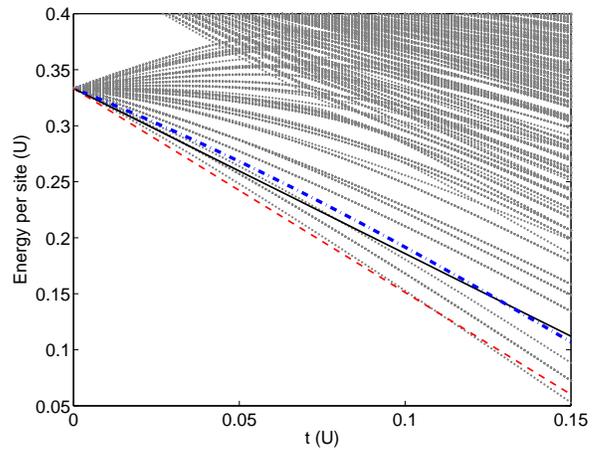} \caption{(Color online) Exact many-body spectrum for 12
particles in a $3\times3$ lattice with $\alpha = 2/3$ considering
only 0,1, and 2 atoms per site (for $\nu = 1/2$, excess particle
density is $\varepsilon = \alpha \nu = 1/3$). Also shown by the
black solid line is our variational estimate of the energy of a
fractional quantum Hall state of excess particles in the presence
of a Mott background.  Dash-dotted blue line shows the Gutzwiller
mean-field superfluid energy for the same density
(1+$\varepsilon$), corresponding to a vortex lattice where the
cores are filled with Mott insulator.  The dashed red line is the
estimate
 of the ground state energy from  Eq.~(\ref{est}), first introduced in \cite{Umucalilar}.
For low enough $t$ the variational energy of the correlated state of excess particles is lower than the superfluid energy. \label{exact3x3}}
\end{figure}

We consider 12 particles in a $3\times 3$ lattice, so that the excess particle density is $1/3$.  We take $\nu=1/2$ and accordingly the number of quanta of flux per plaquette is $\alpha=2/3$.
Fig. \ref{exact3x3} displays the energies (measured in units of $U$) of the first few hundred exact energy eigenstates together with the energies of our two variational wavefunctions: Eqs.~(\ref{FQH ansatz}) and (\ref{gutz}).  We emphasize that our ansatz for the fractional quantum Hall state is not just the Laughlin state, where flux is bound to each particle, but is rather the coexistence of a Mott state and a Laughlin state, with flux bound only to the excess particles.

In Fig. \ref{exact3x3} we also show the estimate from Ref.~\cite{Umucalilar}, which is supposed to describe the correlated state near the Mott insulator,
\be
\label{est}
\Delta E = U n_0 \varepsilon - t (n_0 + 1) f(\alpha)\varepsilon, \ee
where the first term represents the on-site interaction of excess particles with the Mott insulator and the second term is the hopping energy of particles in the Hofstadter ground state denoted by $-t f(\alpha)$, $f(\alpha)>0$ being the dimensionless maximum eigenvalue of the Hofstadter spectrum. Note that $t$ is enhanced by a factor of $(n_0+1)$ due to the Mott background.
No interaction energy is included, as it is expected that in this regime the excess atoms avoid one-another.  It is remarkable how closely this estimate matches the results of the exact diagonalization for small $t$.

For $t \lesssim 0.13$ the energy of our candidate fractional quantum Hall state (with optimized $Z_\beta$) is lower than that of the superfluid, while the opposite holds for larger $t$.  Our physical picture of this is that as $t$ grows the Mott insulator melts, and the density of mobile atoms is no longer commensurate with the magnetic field.

\begin{figure}
\includegraphics[scale=.42]{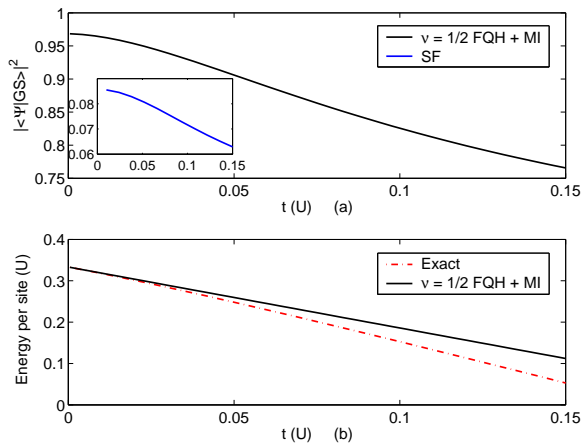}
\caption{(Color online) a) Overlap between the $\nu = 1/2$ FQH + MI state [$|\Psi\rangle$ from Eq.~(\ref{FQH ansatz})] and the exact ground state [$|GS\rangle$, determined from diagonalizing Eq.~(\ref{Bose-Hubbard}) in a truncated basis] as a function of tunneling strength $t$, using the same parameters as Fig.~\ref{exact3x3}. Also shown in the inset is the overlap between a superfluid vortex lattice and $|GS\rangle$. b) Comparison of the variational and exact energies -- from Fig. \ref{exact3x3}.\label{overlap_t}}
\end{figure}

For very small $t$, the fractional quantum Hall state's energy agrees very well with the exact ground state energy: this is shown more clearly in  Fig.~\ref{overlap_t}(b). In Fig.~\ref{overlap_t}(a) we show the overlap between our variational state and the exact ground state. At low $t$ the overlap is greater than 95\%, but it falls off with increasing $t$: presumably due to the increasing importance of particle-hole excitations. The overlap between the ground state and the mean-field superfluid (inset of Fig.~\ref{overlap_t}(a)) is never large, and their energies in Fig.~\ref{exact3x3} never approach one-another. We believe this is due in part to the fact that the mean-field state breaks translational invariance, and consequently involves a superposition of many eigenstates \cite{Mueller}. A quantum superposition of vortex lattices, may in fact be a good alternative description of the fractional quantum Hall state.

Given the small difference between the energies of our two variational states, one must be somewhat cautious about ascribing too much significance to the crossing at $t \sim 0.13$. One also might be concerned that at that value of $t$, both variational states have an energy which is significantly higher than that of the ground state, suggesting that neither may be particularly good descriptions of the true ground state. A third concern is that there is no sign of a phase transition in Fig.~\ref{overlap_t}(a): the overlap between the fractional quantum Hall state and the exact ground state remains above 75\% out to $t\sim 0.15$. Despite these caveats, the large overlaps at small $t$ is convincing evidence that the ground state at low $t$ is a fractional quantum Hall state of excess particles, and it would be surprising if the system formed a correlated state at large $t$.

\subsection{Variational Parameters}

\begin{figure}
\includegraphics[scale=.42]
{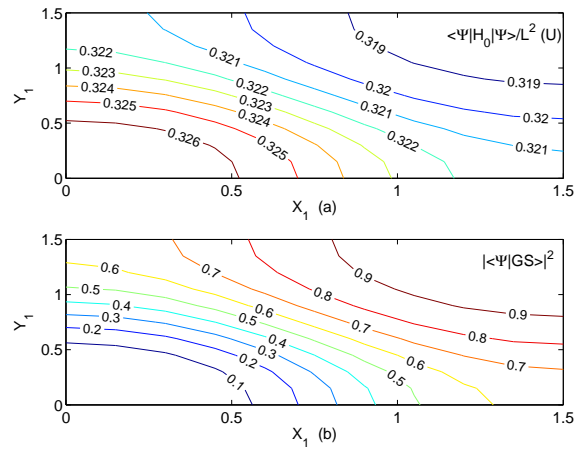}
\caption{(Color online) Variational energy (a) and the overlap with the exact ground state (b) as a function of center of mass zeros $Z_1 = X_1+iY_1$, $Z_2 = L-Z_1$, measured in units of the lattice constant. As with Fig.~\ref{exact3x3}, we consider an $L\times L$ cell with $L = 3$, flux per plaquette $\alpha = 2/3$, filling factor $\nu=1/2$ and total particle number $M=12$. We take $K_x = 0$, $K_y = 0$, and $t = 0.01 U$. The lower the variational energy, the higher the overlap. The lowest energy occurs for $X_1 = Y_1 = L/2$ where the overlap is $96.4\%$. At this point, the variational energy is $0.3186 U$, which is very close to the exact ground state energy of $0.3176 U$.\label{overlap}}
\end{figure}

In Fig.~\ref{overlap} we show how the energy of the variational state depends on the parameters $Z_\beta$, which represent where ``vortices" can be found around which the center of mass flows. The boundary conditions in Eq.~(\ref{boundary conditions}) force the wavefunction to have $m=1/\nu$ of these zeros (in the present case $m=2$). In the absence of the lattice, the energy is invariant under changing these parameters, leading to an $m$-fold degeneracy of the ground state \cite{Haldane2}. Here this symmetry is absent and the energy depends on $Z_\beta$. Not surprisingly, the overlap between the variational state and the exact ground-state is directly correlated with the energy. This overlap has a maximum when the variational energy has a minimum.

\section{Variational Monte-Carlo}\label{monte-carlo}
\begin{figure}
\includegraphics[scale=.42]
{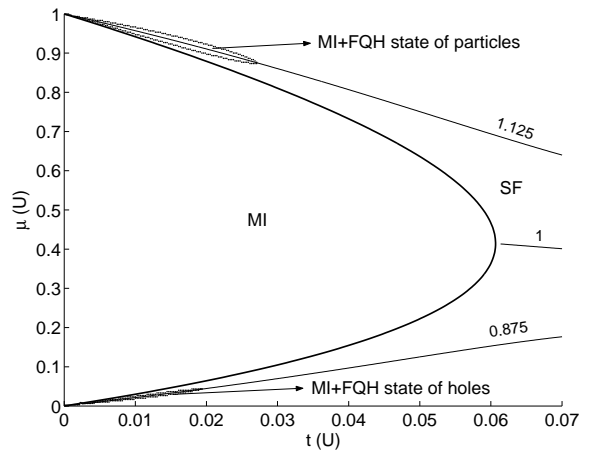}
\caption{Phase diagram for $\alpha = 1/4$ and $\nu = 1/2$. Boundary between Mott insulator (MI) and superfluid (SF) states is found from a mean-field calculation. Excess particle (or hole) density is $\varepsilon = \alpha \nu = 0.125$. Boundary of the coexistent $\nu = 1/2$ FQH state of excess particles (holes) and $n_0 = 1$ MI state centered around the 1.125 (0.875) constant density line is determined from a comparison of VMC and mean-field energies. We consider 8 particles in an $8\times8$ lattice in the VMC calculation. \label{phasediagram}}
\end{figure}

Unfortunately the maximum size of the system which can be treated by the techniques of Sec.~\ref{exact} is quite limited.
Our preceding results for small system size predominantly serve as a guide for physical intuition, and cannot quantitatively describe the physics of the infinite system. Here we introduce a variational Monte-Carlo (VMC) algorithm \cite{Monte Carlo} in order to calculate the energy $\langle \Psi|H|\Psi\rangle=\langle \Psi|H_0-\mu M|\Psi\rangle$, where $M$ is the total number of particles. This will allow us to make a more solid comparison of the energies of the superfluid and correlated states, and draw the phase diagram in Fig.~\ref{phasediagram}. This phase diagram illustrates the regions of  $t-\mu$ plane where either the superfluid or correlated state has a lower energy.

We begin by introducing a basis $|R=\{z_1,\cdots,z_N\}\rangle$ where the $N$ excess particles are at sites $z_1$ through $z_N$. This allows us to write
\begin{eqnarray}\label{VMC expectation}
\langle\Psi| H |\Psi\rangle &=& \sum_{RR^\prime} \langle \Psi|R\rangle \langle R |H|R^\prime\rangle \langle R^\prime |\Psi\rangle
= \sum_R P_R E_R,\nonumber\\
P_R&=&|\langle R|\Psi\rangle|^2,
\\\nonumber
E_R&=&\sum_{R^\prime} \frac{ \langle R |H|R^\prime\rangle  \langle R^\prime |\Psi\rangle}{ \langle R |\Psi\rangle}.
\end{eqnarray}
We use a Metropolis algorithm to sample the sum over $R$.  Starting from some configuration $R_0$ we generate a new one $R_1$ by attempting to move a single particle by one site. We accept the move with probability  $min\{1, P_{R_{1}}/P_{R_{0}}\}$: we then continue the procedure to generate $R_2$, $R_3$, \ldots.  In the resulting Markov chain each configuration $R$ will appear with probability $P_R$. After $S$ steps, the energy is then estimated as $E_S=\sum_{i=1}^S E_{R_i}/S$. As is usual, we discard the first few thousand steps so as not to bias the sum by our choice of initial configuration. We use a binning analysis to estimate the statistical error on our sum \cite{Ambegaokar}.

For each $R$, we calculate $E_R$ directly. The Hamiltonian only connects a finite number of different configurations (those which differ by moving one particle by one site), and the sum is straightforward numerically.

As a further simplification we note that $E(\mu,t)= E_0(\mu)-(1+n_0)t K$, where  $E_0=U n_0 \varepsilon + U (n_0-1)n_0/2 -\mu(n_0+\varepsilon)$ is the expectation value of the on-site terms in $H$ and $-(1+n_0)t K$ is the expectation value of the hopping energy. $K$ is independent of $n_0$, as the only role of the Mott background is to provide a Bose-enhancement term of $(1+n_0)$. By using the Monte-Carlo algorithm to calculate $K$, rather than $E$, we produce $E(\mu,t)$ for all $n_0$, $\mu$, and $t$ at once.

Table~\ref{VMCtable} lists the parameters for which we have
performed VMC calculations. For the smallest system sizes
($L=3,4,5$) we find that the VMC agrees with the direct
calculation of the variational energy.  From the chart, we
conclude that finite size effects are significant in the $L=3$
cases, but for larger $L$ the differences between the energies of
the two systems are within a few percent. We have not extrapolated
to $L=\infty$.

\begin{table}

\begin{tabular}{|c|c|c|c|c|c|}
\hline
$\alpha$&$\varepsilon$&$L$&$N$&$K$&$\delta K$\\
\hline
2/3&1/3&3&3&0.7376&$6\cdot 10^{-4}$\\
&&6&12&0.5187&$4\cdot 10^{-4}$\\
\hline
4/9&2/9&3&2&0.4419&$4\cdot 10^{-4}$\\
&&6&8&0.4455& $2\cdot 10^{-4}$\\
\hline
8/25&4/25&5&4&0.3874&$1\cdot 10^{-4}$\\
&&10&16&0.3873&$5\cdot 10^{-5}$\\
\hline
1/4&1/8&4&2&0.3483&$2\cdot 10^{-4}$\\
&&8&8&0.3375& $4\cdot 10^{-5}$\\
\hline
\end{tabular}

\caption{Results of our variational Monte-Carlo calculation.
$\alpha$ is the number of flux per plaqette, $\varepsilon$ is the
density of excess particles, $L$ is the system size, $N$ is the
number of excess particles, $-(1+n_0)t K$ is the hopping energy
per site, where $t$ is the hopping matrix element and $n_0$ is the
number of particles per site in the underlying Mott state.  $K$ is
dimensionless.  Our estimates of the statistical error in $K$ from
a binning analysis of 80000 samples are given by $\delta K$.}
\label{VMCtable}
\end{table}

Fig.~\ref{phasediagram} illustrates our results for $\alpha=1/4$.
Near the constant density line $n=1+\varepsilon$, there is a region where our variational wavefunction has a lower energy than the Gutzwiller mean field vortex lattice. This corresponds to an incompressible $\nu = 1/2$ bosonic Laughlin state above the $n_0 = 1$ Mott insulator. The same argument can be advanced for holes by just changing the creation operators in Eq. (\ref{FQH ansatz}) with annihilation operators leading to a coexisting Mott insulator and FQH state of holes near the $1-\varepsilon$ line, although it is less visible than in the particle case.

\section{Creation and Observation}

Several labs currently have the technology to create a rotating optical lattice \cite{Williams, Tung}, which can be directly used to create the system described here.  Those experiments still are far from the Mott regime, but they are progressing rapidly.  An alternative approach to implementing Eq.~(\ref{Bose-Hubbard}) is to use a non-rotating lattice, and generate the phases on the hopping matrix elements by some other means \cite{Artificial field}.  The most advanced demonstration of this technique was from Lin et al. \cite{Lin}.

One of the more promissing schemes for experimentally observing
the incompressible states described here is through in-situ
imaging of the density profile of a trapped gas
\cite{Palmer,Umucalilar2}. The fractional quantum Hall states
should appear as extra steps in the density profile near the Mott
insulator plateaus. Moreover, as the magnitude of the effective
magnetic field increases, these steps move in predictable ways:
the density is set by the magnetic flux, and the size of the gap
(hence the spatial size of the plateau) increases with magnetic
field.  One can even imagine that for a fixed flux there will
appear a sequence of FQH states with larger even denominators and
thus with smaller densities all the way up to the MI-SF phase
boundary, however their size will be much smaller and they may not
be discernible at all. Other probes for the FQH  states may be
noise correlations in time-of-flight experiments, measurement of
the Hall conductance for the mass current in a tilted lattice, or
Bragg spectroscopy \cite{Hafezi,Sorensen,Palmer,Bhat, Moller,
Cooper, Umucalilar2}.

A major impediment to observing these states is the need to reduce the temperature to below the scale of the gap, which is a fraction of the hopping matrix element $t$.  Such temperatures are currently hard to reach reliably.

\section{Summary}

In summary, we have predicted that experiments on bosons in rotating lattices (or in lattices with an artificial gauge field) will see a phase where the excitations on top of a Mott insulator form a bosonic fractional quantum Hall state. We base our prediction on a set of variational calculations, supplemented by exact diagonalization of small systems. We find that the MI + $\nu=1/2$ Laughlin state has a lower energy than the Gutzwiller mean-field vortex lattice when the density of excess particles/holes, $\varepsilon=N/L^2$, is chosen appropriately ($\varepsilon=\alpha\nu=\alpha/2$), and the hopping $t$ is sufficiently small compared to the interactions $U$. In this regime we find that the overlap between the exact ground state and the Laughlin state is as large as 96\%, but the overlap with the superfluid is smaller than 10\%. We produced a phase diagram (Fig.~\ref{phasediagram}), illustrating where this novel phase should be found at low temperatures.

\begin{acknowledgements}
R.O.U. is supported by T\"{U}B\.{I}TAK. R.O.U. also wishes to
thank Eliot Kapit for useful discussions and acknowledges the
hospitality of the LASSP, Cornell University, where this work has
been completed. This material is based in part upon work supported
by the National Science Foundation under Grant Number PHY-0758104.
\end{acknowledgements}

\end{document}